\documentclass[12pt]{article}
\usepackage{helvet,times,mathptm}
\usepackage{graphicx}
\begin{document}

\begin{center}
{\noindent
{\bf Full Spin and Spatial Symmetry Adapted Technique for Correlated Electronic
Hamiltonians: Application to an Icosahedral Cluster}}
\end{center}

\begin{center}

{\small Shaon Sahoo$^{a,}$ \footnote[1]{shaon@physics.iisc.ernet.in}
and S. Ramasesha$^{b,}$ \footnote[2]{ramasesh@sscu.iisc.ernet.in}}

\end{center}

{\noindent
{\small
a. Department of Physics, Indian Institute of Science, Bangalore 560012, India.
\\
b. Solid State $\&$ Structural Chemistry Unit, Indian Institute of Science,
Bangalore 560012, India.\\
\vspace{.2cm}
~\\
}}

\begin{abstract}
\noindent
One of the long standing problems in quantum chemistry had been the inability 
to exploit full spatial and spin symmetry of an electronic Hamiltonian  
belonging to a non-Abelian point group. Here we present a general technique 
which can utilize all the symmetries of an electronic (magnetic) Hamiltonian 
to obtain its full eigenvalue spectrum. This is a hybrid method based on 
Valence Bond basis and the basis of constant z-component of the total spin. 
This technique is applicable to systems with any point group symmetry and is 
easy to implement on a computer. We illustrate the power 
of the method by applying it to a model icosahedral half-filled electronic 
system. This model spans a huge Hilbert space (dimension 1,778,966) and in 
the largest non-Abelian point group. The $C_{60}$ molecule has this 
symmetry and hence our calculation throw light on the higher energy excited 
states of the bucky ball. This method can also be utilized to study finite 
temperature properties of strongly correlated systems within an exact 
diagonalization approach.
\end{abstract}

\pagebreak

\section{Introduction}

One of the major goals of the electronic structure theory of molecules is the 
determination of the excited states and their properties. For studying the 
linear and nonlinear optical properties of a system, we need to obtain excited 
states of desired symmetries, while we need the full excitation spectrum to 
study finite temperature properties. A brute force diagonalization of the full 
system Hamiltonian is not feasible even for a moderately sized systems and 
even if we succeed in obtaining all the eigenstates, it is difficult to 
identify them with irreducible representations to which they belong. In cases 
where we can manage to obtain the low-lying eigenstates of the Hamiltonian, we 
may miss the states important for the desired purpose, since in correlated 
many-body Hamiltonians, there can be an unpredictable number of `intruder' 
states between the ground and desired excited state. Utilizing the full spacial 
and spin symmetry (conservation of total spin and $z$-component of total spin) 
allows one to obtain several low-lying eigenstates in each spatial symmetry 
subspace for every total spin value and for many low-temperature static 
properties of a system, this will suffice. For the study of dynamic properties 
as well as finite temperature properties, we need to know the full eigen 
spectrum. Obtaining the full eigen spectrum for a large molecular system is,
however, not feasible by any method at the present time. But, utilization of 
all the symmetries of a Hamiltonian allows extending dynamic and finite 
temperature properties to a slightly larger systems than what is feasible in 
the absence of a symmetry. 

Most electronic structure calculations start with molecular orbitals and 
account for correlation by employing a configuration interaction (CI) approach 
either in a perturbative or a variational scheme. However, even a restricted 
CI approach, involving only frontier orbitals, becomes too difficult to handle 
for large molecules \cite{jensen}. We can circumvent this difficulty by 
resorting to model Hamiltonian. In some molecular systems, it is possible to 
identify a subsystem to which the important electronic excitations are 
confined. In such a situation, it is both advantageous and insightful to deal 
with model electronic Hamiltonians which describe the excitations in the 
subsystem. One such molecular system is the conjugated $\pi$ system. 

The model Hamiltonian for describing conjugated system was 
first introduced by H\"uckel and has mainly served pedagogical purpose in 
understanding the chemistry of conjugated systems \cite{salem}. More realistic 
models which take into account electronic repulsions within the $\pi$ system 
was introduced by Pariser and Parr \cite{pariser} as well as by Pople 
\cite{pople} independently in 1953. This and related models such as the 
Hubbard model \cite{hubbard} have dominated the study of correlated electronic 
systems in chemistry and physics for almost half-a-century. 
These models consist of a one-electron Hamiltonian defined in the basis of site 
orbitals and whose matrix elements are non-vanishing along the diagonal as well 
as between orbitals on chemically bonded sites and a two electron term which is 
approximated within a zero differential overlap (ZDO) scheme 
\cite{springborg,zerner}. The ZDO scheme leads to electron repulsion integrals 
which are diagonal in the atomic orbital basis. The Pariser-Parr-Pople (PPP) 
model Hamiltonian is given by \\

\begin{eqnarray}\label{ppp}
{\hat H_{PPP}}~=~ -\sum_{<ij>,\sigma}t_{ij}(\hat c^{\dagger}_{i\sigma} 
\hat c_{j\sigma} ~+~ H.c.)~+~ \sum_i\frac{U_i}{2}\hat n_i(\hat n_i-1)
~+~\sum_{i>j}V_{ij}(\hat n_i-z_i) (\hat n_j-z_j)
\end{eqnarray}

\noindent Here, the first term of the Hamiltonian is the H\"uckel term with 
$\hat c_{i\sigma}^{~\dagger}$ ($\hat c_{i\sigma}$) creating (annihilating) an 
electron of spin $\sigma$ at the $i^{th}$ site and the summation over bonded 
pair of sites $<ij>$. The second term is the Hubbard term with $U_i$ being the 
on-site repulsion energy for i-th site ($\hat n_i$ is the number operator for 
$i^{th}$ site). The last part is the inter-site interaction term with $V_{ij}$ 
being the density-density electron-repulsion integral between sites $i$ and 
$j$, $z_i$ is the local chemical potential and corresponds to the occupancy of 
$i^{th}$ site for which the site is neutral. We employ the Ohno interpolation 
scheme to parametrize $V_{ij}$ \cite{ohno}.

\begin{eqnarray}\label{Ohno}
V_{ij}=14.397\left[\left(\frac{28.794}{U_i+U_j}\right)^2+r_{ij}^2\right]
^{-1/2}
\end{eqnarray}

\noindent Here $r_{ij}$ is the distance (in \AA\hspace*{.09cm}unit) between 
the $i^{th}$ and $j^{th}$ sites using the Hubbard $U$'s (in eV) at these sites.

The Fock space of the PPP Hamiltonian scales as $4^N$ where $N$ is the number 
of orbitals considered in the system and obtaining even a few exact low-lying 
states of the Hamiltonian for reasonable $N$ could pose a challenge. While this 
problem can be managed to some extent by resorting to approximate treatments 
such as restricted CI schemes by (1) restricting the number of active orbitals 
considered in the CI step and (2) by considering only some classes of 
particle-hole excitations of the system \cite{jensen}, the advantage of 
exploiting all the 
symmetries possessed by the PPP Hamiltonian cannot be overstated. Full 
symmetry adaptation, besides factorizing the Hilbert space and thereby 
reducing computational effort also provides the symmetry labels of the states 
for discerning the state properties. The PPP Hamiltonian, being 
non-relativistic conserves total spin, $S$, as well as z-component of total 
spin, $M_S$ and could possess additional spatial symmetries depending on the 
system in question. The diagonalization of the Hamiltonian can be simplified 
by specializing the basis, in which the matrix representation of the 
Hamiltonian is sought, to the case of fixed total spin and z-component of the 
total spin and a specific irreducible representation of the point group.
 
The conservation of the $\hat S_{tot}^{z}$, the total z-component of spin is 
achieved by choosing from the Fock space, states whose total $M_{S}$ 
corresponds to the desired value. This is trivially possible by choosing a 
spin orbital basis and populating them with electrons to obtain the desired 
total $M_S$. It is also quite straightforward to set up the Hamiltonian matrix 
in this basis and solve for a few low-lying states in cases where the Hilbert 
space is spanned by a few hundred million states (see subsection 2.1). 
Factorizing the Hilbert space into different irreducible representations of 
the point group of the Hamiltonian is also straightforward as the resultant of 
a spatial symmetry operator, operating on a Slater determinant is easy to 
obtain in atomic orbital basis. In modern quantum chemical calculations, these 
symmetries are routinely employed.

However, construction of spin adapted configuration state functions which are 
simultaneous eigenstates of $\hat S_{tot}^2$ and $\hat S_{tot}^z$ operators is 
nontrivial and pursuit of this has been a long standing interest in quantum 
chemistry. The Hamiltonian matrix in such a symmetrized basis leads to 
matrices of smaller order besides allowing automatic labeling of the states by 
the total spin. Furthermore, the eigenvalue spectrum is enriched, since we can 
obtain several low-lying states in each total spin sector. This can be 
contrasted with obtaining several low-lying states in a given total M$_S$ 
sector which would have states with total spin S$_{tot}\geq M_S$. There are 
many ways of achieving this task \cite{pauncz}; most important among these are 
Valence Bond (VB) approach \cite{soos}, L$\ddot{o}$wdin spin projection 
technique \cite{lowdin,bernu} and group theoretical approaches 
\cite{saxe,duch}. While they are satisfactory regarding spin 
adaptation, most of these techniques virtually fail while dealing with 
non-Abelian spatial symmetry. They become symmetry-specific, even frequently 
impractical while applied to large system with a non-Abelian symmetry (see 
review in \cite{sahoo}). Here we present our hybrid VB-constant $M_S$ method, 
which overcomes these difficulties.

The ultimate goal of symmetry adaptation is to exploit the full spatial and 
spin symmetries of the system, both for computational efficiency and for 
complete labeling of an eigenstate by the total spin and the irreducible 
representation it which belongs. In Sec. 2, we present our hybrid VB-constant 
$M_S$ method which allows exploiting the full spin and spatial symmetries of 
any arbitrary point group. Similar method applicable only to pure spin systems 
has recently been developed \cite{sahoo}. The technique presented here is 
applicable to more general systems of correlated electrons. In Sec. 3, we 
illustrate an application of this method to a PPP and Hubbard model of the 
half-filled icosahedron which has one orbital at each of 
the 12 vertexes. The icosahedron is the smallest system with all the 
symmetries of $C_{60}$, the carbon Bucky ball and obtaining all the 
eigenstates of this model will throw light on the correlated states of 
$C_{60}$.  In Sec. 4, we summarize and  discuss the technique.
  
\section{Hybrid VB and Constant {\it M$_{S}$} Basis Method} 

\begin{figure}[t]
\begin{center}
\hspace*{1cm}{\includegraphics[width=12.0cm]{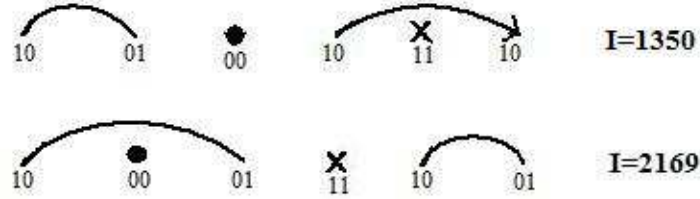}}
\caption{{\small Representation of VB diagrams for a half-filled 6 orbital 
(site) system. Here $\bullet$ denotes empty site, $\times$ denotes doubly 
occupied site. The top VB diagram shows a spin pairings to yield a state with 
total spin $S_{tot}$=1,  its bit representation corresponds to a unique 
integer I = 2350. The bottom VB diagram shows a $S_{tot}$ = 0 state, the 
corresponding unique integer, I is 2169.}}
\label{fig1}
\end{center}
\end{figure}

In an electronic system, a given orbital can be in one of four states; it can 
be (i) empty, (ii)singly occupied with an up spin electron, (iii) singly 
occupied with a down spin electron  and (iv) can be doubly occupied. Constant 
$M_S$ bases, for a given filling of the orbitals, are obtained trivially by 
choosing states from Fock space, whose total $S_z$ value corresponds to the 
desired $M_S$ value ($M_S$ = sum of z-components of individual electron-spins). 
By construction they are orthonormal. The easiest way of constructing the spin 
adapted functions is the diagrammatic valence bond (VB) method based on 
Rumer-Pauling rules \cite{pauling,soos}. If $N$ is the number of orbitals, 
$N_e$ is the number of electrons with $N_{\uparrow}$ up-spin electrons and 
$N_{\downarrow}$ down-spin electrons ($N_e = N_{\uparrow} + N_{\downarrow}$), 
then, all possible linearly independent and complete set of states with total 
spin S and $M_S = S$, for a fixed occupancy of the orbitals, according to 
extended Rumer-Pauling rules are obtained as follows. (i) The N orbitals are 
arranged as dots on a straight line. (i) Doubly occupied sites are marked as 
crosses. (ii) An arrow is passed through 2S of the singly occupied vertexes, 
passing on or above the straight line on which the system is represented. The 
arrow denotes the spin coupling corresponding to total spin $S$ and total 
z-component $M_S = S$. (iii) Remaining singly occupied vertexes are singlet 
paired and are denoted by  lines drawn between them which lie on or above the 
straight line describing the system. (iv) Diagrams with (a) two or more 
crossing lines or (b) crossing line and the arrow or (c) a line enclosing the 
arrow are rejected. The remaining set of diagrams correspond to a complete and 
linearly independent set of VB states for the chosen orbital occupancy. The 
set of VB diagrams which obey the extended Rumer-Pauling rules would hence 
forth be called $``legal"$ VB diagrams. Some legal VB diagrams are shown in 
Fig. (\ref{fig1}) along with the integers which represent them. In the case of 
$N_e$ odd and $S = 1/2$, we cannot have an arrow with just one site! We handle 
this situation by augmenting the system by adding a $``phantom"$ site. The VB 
states of all legal singlets with single occupancy of the phantom site 
provides the complete and linearly independent basis. The phantom site appears 
only in the basis and not in the system Hamiltonian. 

We can generate the complete set of VB states for our case of $N$ orbitals 
with $N_e$ electrons of total spin $S$ and z-component of total spin $M_S = S$ 
by exhausting all possible occupancies of orbitals which satisfy 
$2S = N_{\uparrow} - N_{\downarrow}$. Since an orbital can be in any one of 
four states (empty, doubly occupied, a singlet line beginning or a singlet line 
ending, sites involved in an arrow being treated as line beginnings) we can 
use two bits to represent the state of an orbital. Thus, each VB diagram can 
be uniquely represented as an integer on a computer. 

A line in the VB diagram, between sites $``i"$ and $``j"$, represents ${(\hat 
a^\dagger _{i,\alpha} \hat a^\dagger_{j,\beta}-\hat a^\dagger_{i,\beta}\hat 
a^\dagger_{j,\alpha})|0>} $/$\sqrt{2}$, where we choose $\alpha$ to 
correspond to $|\uparrow \rangle$ and $\beta$ to $|\downarrow \rangle$ 
orientations of the electron. The doubly occupied site  $``i"$ corresponds to 
the state $a^\dagger _{i,\alpha} \hat a^\dagger_{i,\beta}|0>$. The phase 
convention assumed for a line between sites $``i"$ and $``j"$ is that the 
ordinal number $``i"$ is less than the ordinal number $``j"$. The $2S$ singly 
occupied sites  $k_{1}$ $k_{2}$ $k_{3}$ . . . . $k_{2S}$ in the arrow 
represent the state with $M_S = S$ given by $\hat a^\dagger _{k_1,\alpha}\hat 
a^\dagger_{k_2,\alpha} \hat a^\dagger_{k_3,\alpha} . . .\hat a^\dagger_{k_{2S},
\alpha}|0>$. VB states corresponding to other M$_S$ value for this state with 
spin S, can be obtained by operating, required number of times by the 
S$_{tot}^{-}$ operator on the state. Since Hamiltonian in Eq.(\ref{ppp}) is 
isotropic, each eigenstate in the spin S sector is (2S+1) fold degenerate and, 
by Wigner-Eckart theorem \cite{sakurai} it is sufficient to work in subspace of 
chosen M$_S$ value. The VB state corresponding to a given diagram is a product 
of the states representing the constituent parts of the diagram, in no 
particular order as each part is either a product of two Fermion operators or 
a linear combination of the product of two Fermion operators.

Given the definition of a line in the VB diagram, every VB diagram, $|\psi_i>$ 
can be broken up into a linear combination of the constant $M_{S}$ basis 
states \{$|\phi_j>$\} as,

\begin{eqnarray}\label{decomp}
|\psi_i> = \sum_j C_{ij}|\phi_j>
\end{eqnarray}

A VB diagram with $n$ singlet lines yields $2^{n}$ basis states in the 
constant $M_{S}$ basis. To effect the conversion of VB diagrams to constant 
$M_{S}$ functions, we note that each singlet line gives two states; in one 
state, the site at which a singlet line begins is replaced by an $\alpha$ spin 
while the one at which it ends by a $\beta$ spin with phase +1 and in the 
other the spins are reversed and the associated phase is -1. There is a 
normalization constant, $(2^{-n/2}$, associated with the constant $M_{S}$ 
basis state. The matrix relating the VB basis states to constant $M_{S}$ basis 
states, {\bf C}, is a $V\times M$ matrix, where $V$ is the dimensionality of 
the VB space and $M$ that of the constant $M_{S}$ space. If {\bf R$_M$} the 
matrix representation of symmetry operation $\hat R$ is known, in constant 
$M_{S}$ basis, then the knowledge of the matrices {\bf C} and {\bf R$_M$} 
gives the result of operating by the symmetry operator $\hat R$ on a VB state 
as a linear combination of the constant $M_S$ basis states via the matrix 
{\bf B$_{\hat R}$ = CR$_M$}. The projection operator for projecting out the 
basis states on to a chosen irreducible representation $\Gamma$ of the point 
group is given by,

\begin{eqnarray}\label{project}
\hat P_{\Gamma}=\sum_{\hat R}\chi_{\Gamma}^{irr}({\hat R}){\hat R}
\end{eqnarray}

\noindent where, $\chi_{\Gamma}^{irr}(\hat R)$ is the character under the 
symmetry operation $\hat R$ of the point group of the system \cite{bishop}. 
The matrix representation of $\hat P_{\Gamma}$ in the mixed VB and constant 
$M_S$ basis is given by,

\begin{eqnarray}\label{Qgamma}
{\mathbf Q_{\Gamma}}=\sum_{\hat R}\chi_{\Gamma}^{irr}(\hat R)
{\mathbf B_{\hat R}}
\end{eqnarray}

\noindent where, {\bf Q$_{\Gamma}$} is a $V \times M$ matrix. However, the 
rows of the matrix {\bf Q$_{\Gamma}$} are not linearly independent, since the 
complete symmetrized basis transforming as $\Gamma$ spans a much smaller 
dimensional Hilbert space. The exact dimension V$_\Gamma$ of the Hilbert space 
spanned by the system in the irreducible representation $\Gamma$ can be known 
{\it a priori} and is given by,

\begin{eqnarray}\label{Vgamma}
V_{\Gamma}=(d_\Gamma /h)
\sum_{\hat R}\chi^{red}(\hat R)\chi_{\Gamma}^{irr} (\hat R)
\end{eqnarray}

\noindent where $d_\Gamma$ is the dimensionality of the irreducible 
representation $\Gamma$, $h$ is the number of symmetry elements in the point 
group and $\chi^{red}(\hat R)$ is the reducible character for the operation 
$\hat R$. The determination of $\chi^{red}(\hat R)$ is nontrivial and 
the method of computing it will be discussed in the next subsection. The 
$V_{\Gamma}\times M$ projection matrix, {\bf P$_{\Gamma}$} of rank 
$V_{\Gamma}$ is obtained by Gramm-Schmidt orthonormalization of the rows of 
the matrix {\bf Q$_{\Gamma}$} until $V_{\Gamma}$ orthonormal rows are 
obtained. These orthonormal and linearly independent rows yield the desired 
linear combinations which transform as $\Gamma$ and also have total spin $S$. 
Projection matrix {\bf P$_\Gamma$} is represented by these $V_\Gamma$ rows.

The $M\times M$ Hamiltonian matrix {\bf H$_M$} is constructed in the constant 
$M_S$ basis (see subsection 2.1). Since the basis states in this 
representation are orthonormal, we do not encounter the problem of 
$``illegal"$ VB states. In the pure VB method, the Hamiltonian operating on a 
legal VB state can yield illegal VB diagrams which then need to be 
re-expressed as linear combination of the legal VB functions \cite{ramasesh}. 
The $V_ {\Gamma}\times V_{\Gamma}$ Hamiltonian matrix in the fully symmetrized 
basis is given by {\bf P$_{\Gamma}$}{\bf H$_{M}$} ${\bf P}_{\Gamma}^{\dagger}$ 
and one could use any of the well known full diagonalization routines to 
obtain the full eigenspectrum or use the Rettrup modification of Davidson 
algorithm \cite{david} to get a few low-lying states of the 
symmetrized block Hamiltonian in the chosen spin and symmetry subspace.

For degenerate irreducible representations, such as the E, T, G or H 
representations, the above procedure does not lead to the smallest block of 
the Hamiltonian matrix. In such cases, it is advantageous to work 
with bases that transform according to one of the components of the 
irreducible representation. In case of E, T or H, this can be achieved by 
choosing an axis of quantization and projecting out basis states of the 
irreducible representation which are diagonal about a rotation about the 
quantization axes. For example, in the case of the irreducible representation 
that transforms as T, we can choose one of the C$_{3}$ axes as a quantization 
axis and project the basis states which transform as the irreducible 
representation T, using $(I+C_{3}^{1}+C_{3}^{2})$ as the projection operator. 
This operator projects states that transform as the Y$_{1}^{0}$ component of 
the three fold degenerate irreducible tensor operator. Similarly we can choose 
a $C_5$ axis and use $(I+C_5^1+C_5^2+C_5^3+C_5^4)$ as projection operator 
for the irreducible representation H. For the E representation, we can use 
$(I+C_2)$ as projection operator with a chosen $C_2$ axis. The case of G is a 
bit tricky; one has to choose two $C_2$ axes, orthogonal to each other. The 
projection operator for this case then would be: $(I+C_2)(I+C'_2)$. After 
these projection operations, dimensions of the Hamiltonians to be diagonalized 
would be half, one third, one fourth or one fifth respectively for the E, T, G 
and H representations.

Here we wish to emphasize the computational advantage of our technique over 
the  constant $M_{S}$ basis method. The additional steps involved in the hybrid
VB-Constant M$_S$ method are (i) construction of the {\bf C} matrix and (ii) 
computation of the {\bf B$_{\hat R}$} matrix. However, if we wish to compute 
the properties of a state expressed as a linear combination of VB diagrams, 
the simplest way is to use the {\bf C} matrix to transform the state from the 
VB basis to the constant $M_S$ basis. Therefore, construction of the {\bf C} 
matrix is not strictly an overhead. Besides,the construction of the {\bf C} 
matrix is a very fast step as the row index of the {\bf C} matrix is the index 
of the VB state which we wish to decompose and the column indices of elements 
in this row are the indices of the resultant constant M$_S$ states. The 
constant M$_S$ states are easily generated as an ordered sequence of integers 
which represent them and this facilitates searching for the column index of 
the matrix. The coefficients will have a magnitude of $2^{-n/2}$ where $n$ is 
the number of lines in the $i^{th}$ VB diagram; the phase of the coefficient 
is easily fixed based on the phase convention used for a singlet line. In the 
hybrid approach, computation of the {\bf B$_{\hat R}$} matrix involves the 
matrix multiplication, {\bf C}{\bf R$_M$}.
The number of arithmetic operations involved is however very small, since both 
{\bf C} and {\bf R$_M$} are sparse matrices with the latter having only one 
nonzero matrix element per row. In both constant M$_S$ and hybrid approaches 
one has to obtain the projection matrix {\bf P$_{\Gamma}$} by retaining only 
the orthogonal rows of the matrix {\bf Q$_\Gamma$}. Since the number of 
orthogonal rows in {\bf Q$_{\Gamma}$} is far fewer than in {\bf R$_M$}, this 
step is faster in the hybrid approach than in the constant $M_S$ approach by a 
factor D($\Gamma_S$)/D($\Gamma_{M_S}$), where D($\Gamma_S$) is the 
dimensionality of the space of the irreducible representation $\Gamma$ with 
spin S and D($\Gamma_{M_S}$) is similarly the dimension of the space $\Gamma$ 
with constant M$_S$. Though, this advantage is largely off-set by the fact 
that the {\bf R$_M$} matrix in constant $M_S$ basis is more sparse than the 
{\bf Q$_{\Gamma}$} matrix in the hybrid approach. Computation of the 
eigenvalues (diagonalization of Hamiltonian) in the constant $M_S$ approach is 
slower than in the hybrid approach, since D($\Gamma_{M_S}$)$>$D($\Gamma_S$) 
for most S (for example see Table \ref{tbl1}). The memory required for the 
hybrid approach is not very different from that of constant $M_S$ approach, 
even though the matrices in the hybrid approach are slightly denser, they are 
smaller in size. The only additional memory demand  in the hybrid approach is 
the storage of sparse {\bf C} matrix. The major advantage of the hybrid 
approach is that we can obtain a far richer spectrum, since we are targeting 
each spin sector separately, unlike in the constant $M_S$ approach. Thus, if 
we can obtain (by our approach), say 10 states in each S sector, then each one 
will correspond to a unique state. There will be no repetition of the states. 
But in contrast, 10 states obtained in a $M_S$ sector (by constant $M_S$ 
approach) may not be unique, since many of these states would be repeated in 
different $M_S$ sectors.

\subsection{Implementation Details}

We can represent a basis uniquely by a 2N-digit binary number \cite{soos} (N 
is the number of sites / orbitals); the first two bits describe the state of 
the first site, next two bits describe the state of the second and so on. For 
constant $M_S$ basis, we use the bit states ``00" for an empty site, ``10" for 
site with spin-up electron, ``01" for site with spin-down electron and ``11" 
for a doubly occupied site. Similarly, for VB basis, we use ``00" for empty 
site, ``10" for a singlet line-beginning at a site as well as for all sites in 
the arrow, ``01" for a line-ending and ``11" for a doubly occupied site. The 
Rumer-Pauling rules are implemented by enforcing that for a given site $``i"$, 
the quantity (\# of line beginnings - \# of line endings) at sites one to 
$i-1$ should be $\geq 0$ \cite{ramasesh1}. Our binary coding implies that the 
number of bits in the state ``1" is equal to the number of electrons $N_e$. To 
generate integers that represent VB states, we generate integers with $N_e$ 
``1" bits in the bit field from zero to (2N-1) in increasing sequence and 
check to see if the bit pattern corresponds to a 
``legal" VB diagram with chosen total spin.  For integers corresponding to 
constant $M_S$ basis, total $M_S$ value should be the desired value and the 
Rumer-Pauling condition is not enforced. The positive integers so generated 
uniquely represent the states of the VB or constant $M_S$ bases.  

Computationally, finding the transformation matrix {\bf C} which carries the 
VB basis to constant $M_{S}$ basis is straightforward. We initialize the 
coefficients in the row of the matrix {\bf C} corresponding to the chosen VB 
state to zero. We then decompose the VB diagram by converting every singlet 
line in the diagram into two $M_S$ states. The indices of the resulting 
constant $M_{S}$ states correspond to the column indices of {\bf C} and are 
determined by a binary search on the list of integers that represent the 
constant $M_{S}$ states. The corresponding matrix element is given by the 
normalized VB coefficient with appropriate phase. On a computer, the 
transformation matrix {\bf C} is stored in sparse form.  Next, we construct 
the projection matrix ({\bf P$_{\Gamma}$}), by constructing the matrix 
representation of each of the symmetry operators, $\hat R$, of the point group 
in the constant $M_{S}$ basis. This is achieved by (i) obtaining the 
occupancies of each site from the integer representing the basis state, and 
(ii) by letting $\hat R$ act on the basis state by appropriately rearranging 
the sites together with their occupancies to obtain the new bit pattern 
corresponding to the resultant state. The new occupancy pattern is converted 
into the integer representing the state and fixing the column index of the 
matrix {\bf R} by a binary search for 
the index of the new integer in the list of integers representing the constant 
$M_S$ basis. Care should be taken to keep track of the phase factor while 
interchanging the occupancies since Fermion creation operators anticommute. 
From a knowledge of {\bf C} and all the {\bf R$_M$} matrices we can construct 
the matrix ${\mathbf Q_{\Gamma}}$ (Eq. \ref{Qgamma}). But the rows of 
${\mathbf Q_{\Gamma}}$ are in general not linearly independent, eliminating 
linear dependencies leads to the projection matrix {\bf P$_{\Gamma}$} with 
V$_{\Gamma}$ number of linearly independent rows.  The V$_{\Gamma}$ linearly 
independent rows can be obtained by (i) collecting all linearly independent 
rows, by inspection, by noting that the set of rows which are disjoint (that is 
do not have non zero elements with common column index) are orthogonal by 
virtue of the fact that the constant $M_S$ basis sets are orthogonal and (ii) 
by carrying out Gram-Schmidt 
orthonormalization to obtain the remaining linearly independent rows. However, 
knowing V$_{\Gamma}$ {\it a priori} is important to be able to stop the 
orthonormalization process once the number of linearly independent rows 
obtained equals the dimensionality of the symmetrized space. While 
V$_{\Gamma}$ can be obtained from Eq. \ref{Vgamma}, it needs a knowledge of 
the reducible character.

To obtain the reducible character, it appears that we need a matrix 
representation of the symmetry operator in the VB basis. Given an operator 
$\hat R$, the matrix representation {\bf r}, in the VB basis, is obtained from,

\begin{eqnarray}\label{RVB}
\hat R |\psi_i>  = \sum_j r_{ij} |\psi_j>.
\end{eqnarray}

However, since the VB basis is non-orthogonal, we need the inverse of the 
overlap matrix,{\bf S$^{-1}$}, where S$_{ij} = <\psi_i|\psi_j>$ are the matrix 
elements of {\bf S}. The matrix {\bf r} is then given by {\bf RS$^{-1}$}, 
where the matrix elements of {\bf R} are given by $R_{ij} = <\psi_j|\hat R|
\psi_i>$. In general determination of the matrix {\bf S$^{-1}$} is difficult 
for Fermionic systems and computationally prohibitive for large pure spin 
systems.

The above difficulty can be circumvented by resorting to the bit 
representation of VB and constant $M_S$ basis states. Using the {\bf C} 
matrix, we can rewrite \ref{RVB} as 

\begin{eqnarray}\label{RMS}
\hat R |\psi_i>  = \sum_j r_{ij} \sum_k C_{jk} |\phi_k>.
\end{eqnarray}

For every state $|\psi_i>$, we need to find the coefficient $r_{ii}$ and the 
reducible character $\chi^{red}(\hat R) = \sum_i r_{ii}$. Taking the inner 
product on both sides of Eq. \ref{RMS} with $|\phi_l>$,we get,

\begin{eqnarray}\label{matelemrhs}
<\phi_l|\hat R |\psi_i>  = \sum_j \sum_k r_{ij} C_{jk} <\phi_l|\phi_k> ~~~~~~~
\\ \nonumber
=\sum_j r_{ij} C_{jl}~~~~~~~~~~~~~~~~~~~~~~~~~~~~~
\end{eqnarray}

$r_{ij}$ are unknowns and need to be determined. The $lhs$ can also be 
evaluated as

\begin{eqnarray}\label{matelemlhs}
<\phi_l|\hat R |\psi_i>=<\phi_l|\hat R|\sum_j C_{ij}|\phi_j> ~~~~~~~~
\\ \nonumber
=\sum_j C_{ij} \sum_k R_{jk} <\phi_l|\phi_j> \\ \nonumber
=\sum_k R_{lk} C_{il}~~~~~~~~~~~~~~~~~~~~~~~
\end{eqnarray}

\noindent where $R_{ij}$ is the matrix representation of $\hat R$ in the 
constant $M_S$ basis which is known. The only unknowns on the $rhs$ of 
Eq. \ref{matelemrhs} are the coefficients $r_{ij}$ and we need to determine 
the diagonal elements $r_{ii}$. 

To determine $r_{ii}$, let us first assume that integers $\{J_l\}$ represent 
the constant $M_S$ basis states $\{|\phi_l>\}$ and the integers $\{I_i\}$ 
represent the VB states $\{|\psi_i>\}$. Now we note that in the expansion of a 
VB state $|\psi_i>$ as a linear combination of constant $M_S$ states 
(Eq. \ref{decomp}), the largest integer that represents the constant $M_S$ 
state, $J_l$  which appears in the expansion, is the one corresponding to the 
integer that represents the VB state $|\psi_i>$ itself, namely $I_i$. This is 
because, we have chosen the bit state ``10" both for a line beginning in the 
VB state and for an up spin occupancy in the constant $M_S$ basis. The 
assertion that $I_i \geq J_j$ in the decomposition of the VB state $|\psi_i>$ 
into constant $M_S$ functions $|\phi_j>$, implies that the matrix {\bf C} has 
nonzero elements $C_{jk}$ only for $J_k \leq I_j$. 

Let us consider Eq. \ref{RVB}, we note that on the {\it rhs} the summation runs 
over all the states of the VB basis. Let us consider the VB state, $|\psi_V>$ 
which is represented by the largest permitted integer, $I_{V}$. This integer 
also correspond to the $M_S$ basis state $|\phi_M>$ (where $M$ is the 
dimensionality of the constant $M_S$ space) with the largest integer 
representation, ($I_{V} = J_{M})$, Taking the inner product with the state 
$|\phi_{M}>$, from Eq. \ref{matelemlhs} and Eq. \ref{matelemrhs}, we obtain,

\begin{eqnarray}\label{riNv}
<\phi_{M}|\hat R|\psi_i> = r_{i,V} C_{V,M}
\end{eqnarray}

All other terms on the {\it rhs} of Eq. \ref{matelemrhs} are zero. Hence using 
Eq. \ref{riNv}, we can determine $r_{i,V}$.  We can now proceed with the 
constant $M_S$ state whose representing integer $J_{K}$ is equal to 
$I_{V-1}$. The constant $M_S$ state with $J_{K}$ can appear only in the 
expansions of the VB states $\psi_{V}$ and $\psi_{V-1}$. Taking the inner 
product with $\phi_{K}$, we obtain,

\begin{eqnarray}\label{riNvm1}
<\phi_{K}|\hat R |\psi_i> = r_{i,V}C_{V,K} + r_{i,V-1}C_{V-1,K}
\end{eqnarray}

In \ref{riNvm1}, the only unknown is $r_{i,V-1}$ and can be evaluated. 
Similarly, by proceeding to the VB state $\psi_{V-2}$, we can obtain 
$r_{i,V-2}$. We can terminate when we reach the VB state $|\psi_i>$. This 
procedure can be adopted to obtain all the diagonal elements of the {\bf r} 
matrix and hence the reducible character, $\chi^{red}$.

Constructing the Hamiltonian matrix in constant $M_S$ basis in real space is a 
fast and easy step. The basis states are eigenstates of the interaction part 
for the model Hamiltonians. From the binary sequence of the integers which 
represent the constant $M_S$ basis, we know the occupancy of each site and 
hence the diagonal contribution of the interaction terms. Simple rules for 
operating on a constant $M_S$ basis state by the operators $\hat E_{ij} = 
\sum_\sigma (\hat a^\dagger_{i \sigma} \hat a _{j \sigma} + \hat a^\dagger_{j 
\sigma} \hat a _{i \sigma})$ have been published elsewhere and together with a 
binary search procedure which allows rapid generation of the matrix 
corresponding to the one-electron terms of the Hamiltonian \cite{soos}. 

\section{Application to Icosahedral Cluster}

\begin{figure}[t]
\begin{center}
\hspace*{1cm}{\includegraphics[width=12.0cm]{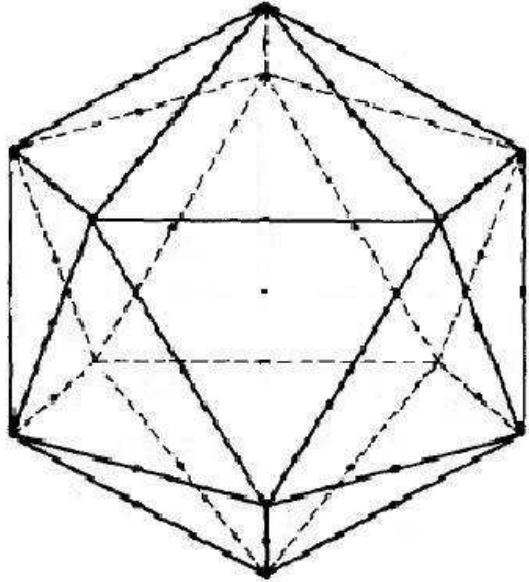}}
\caption{{\small Regular icosahedron with ($I_h$) symmetry. Our system has an
orbital at each of the 12 vertexes, and a transfer integral corresponding to a
bond on each of the 30 edges; In the PPP model, the bond length is taken to be 
1.4 \AA.}}
\label{fig2}
\end{center}
\end{figure}

To illustrate the power of our technique, we have applied the method to a 
12-site regular icosahedral cluster (see Fig. \ref{fig2}) at half-filling. It 
has 30 edges (each one here taken to be of length 1.4 \AA) representing a 
chemical bond.We have chosen this system, because it belongs to very high 
symmetry non-Abelian point group and presents a very general case for testing 
our method. This point group is also the same as the point group of the 
$C_{60}$ molecule. So properties, which are particularly symmetry-related, 
obtained for our model system, would also be useful in gaining insights into 
the $C_{60}$ molecule. 

We have studied the icosahedron within the Hubbard model in which the 
inter-site interactions are neglected, for a range of $U$ values, as well as 
in the PPP model with standard Carbon parameters. The number of $M_S$=0 states 
is 853,776 and we have obtained the exact energies of all the states, by using 
the full spatial and spin symmetries of the system. The energies of 
$M_S \neq 0$ states are also known, since we know the total spin of each 
state. We have studied the density of states in each symmetry and spin sector 
as a function of $U$ in the Hubbard models and also for standard parameters in 
the PPP model. 

In Table (\ref{tbl1}), we give the dimensions of all the subspaces of 
different  total spin and total $M_S$ values, for the Icosahedral cluster. 

\begin{table}[htbp]
\caption{{\small Dimensionalities of different spin subspaces of a half-filled
12-site (orbital) electronic system. $D(S)$ is the dimensionality of the
constant S basis and $D(M_S)$ is the dimensionality of the constant $M_S$
basis.}}
\begin{center}
\begin{tabular}{|l||r|r|r|r|r|r|r|}
\hline
S/M$_{S}$ & 0 & 1 & 2 & 3 & 4 & 5 & 6 \\ \hline
$D(S)$ & 226512 & 382239 & 196625 & 44044 & 4212 & 143 & 1 \\ \hline
$D(M_{S})$ & 853776 & 627264 & 245025 & 48400 & 4356 & 144 & 1 \\ \hline
\end{tabular}
\end{center}
\label{tbl1}
\end{table}

We note here the huge fall in size of total spin subspaces compared to total
$M_S$ subspaces for most of the cases. Using the hybrid VB-constant $M_S$
method, we have broken down each total spin sectors into basis states that
transform as different irreducible representations of the icosahedral point
group. The dimensionalities of the various symmetry subspaces are shown in
Table (\ref{tbl2}).

\begin{table}[h]
\caption{{\small Dimensionalities of different symmetry and spin subspaces of
half-filled icosahedral cluster.}}
\begin{center}
\begin{tabular}{|l||r|r|r|r|r|r|r|} \hline
S$_{tot} \rightarrow$ & & & & & & & \\
$\Gamma$ $\downarrow$ & 0 & 1 & 2 & 3 & 4 & 5 & 6 \\ \hline \hline
A$_{g}$ & 2040 & 3128 & 1684 & 382 & 38 & 3 & 1 \\ \hline
T$_{1g}$ & 16602 & 28821 & 14625 & 3261 & 309 & 6 & 0 \\ \hline
T$_{2g}$ & 16602 & 28821 & 14625 & 3261 & 309 & 6 & 0 \\ \hline
G$_{g}$ & 30272 & 50932 & 26236 & 5880 & 568 & 16 & 0 \\ \hline
H$_{g}$ & 47940 & 79305 & 41255 & 9220 & 900 & 40 & 0 \\ \hline
A$_{u}$ & 1852 & 3188 & 1644 & 348 & 40 & 0 & 0 \\ \hline
T$_{1u}$ & 17082 & 28686 & 14700 & 3372 & 294 & 18 & 0 \\ \hline
T$_{2u}$ & 17082 & 28686 & 14700 & 3372 & 294 & 18 & 0 \\ \hline
G$_{u}$ & 30160 & 50992 & 26176 & 5888 & 560 & 16 & 0 \\ \hline
H$_{u}$ & 46880 & 79680 & 40980 & 9060 & 900 & 20 & 0 \\ \hline \hline
Tot Dim $\rightarrow$ & 226512 & 382239 & 196625 & 44044 & 4212 & 143 & 1 \\
\hline
\end{tabular}
\end{center}
\label{tbl2}
\end{table}

We note that most subspaces are small enough for obtaining all the eigenstates
of the Hamiltonian. However, for degenerate representations the subspaces are 
large and can be reduced by a factor equal to the dimensionality of the 
representation, as described earlier. We have used this approach and obtained 
all the eigenstates of the ensuing Hamiltonian matrix using a full matrix 
diagonalization routine. Since the number of eigenstates in each subspace is 
large, we have computed the density of states (DoS) using a $\Delta$E of 
0.4eV, for which the histograms of the DoS are stable. A histogram for 
particular spin is evaluated by summing over corresponding states of all 
symmetries. Same is for histogram for particular symmetry, where corresponding 
states of all spins are considered. Although, unlike the 
one-particle DoS, the many-body DoS is not an intensive quantity. However, for 
a given model and system size, we can use these quantities to understand the 
behavior of the system.

\subsection{Hubbard model studies}

\begin{figure}[h]
{\includegraphics[width=1.0\textwidth]{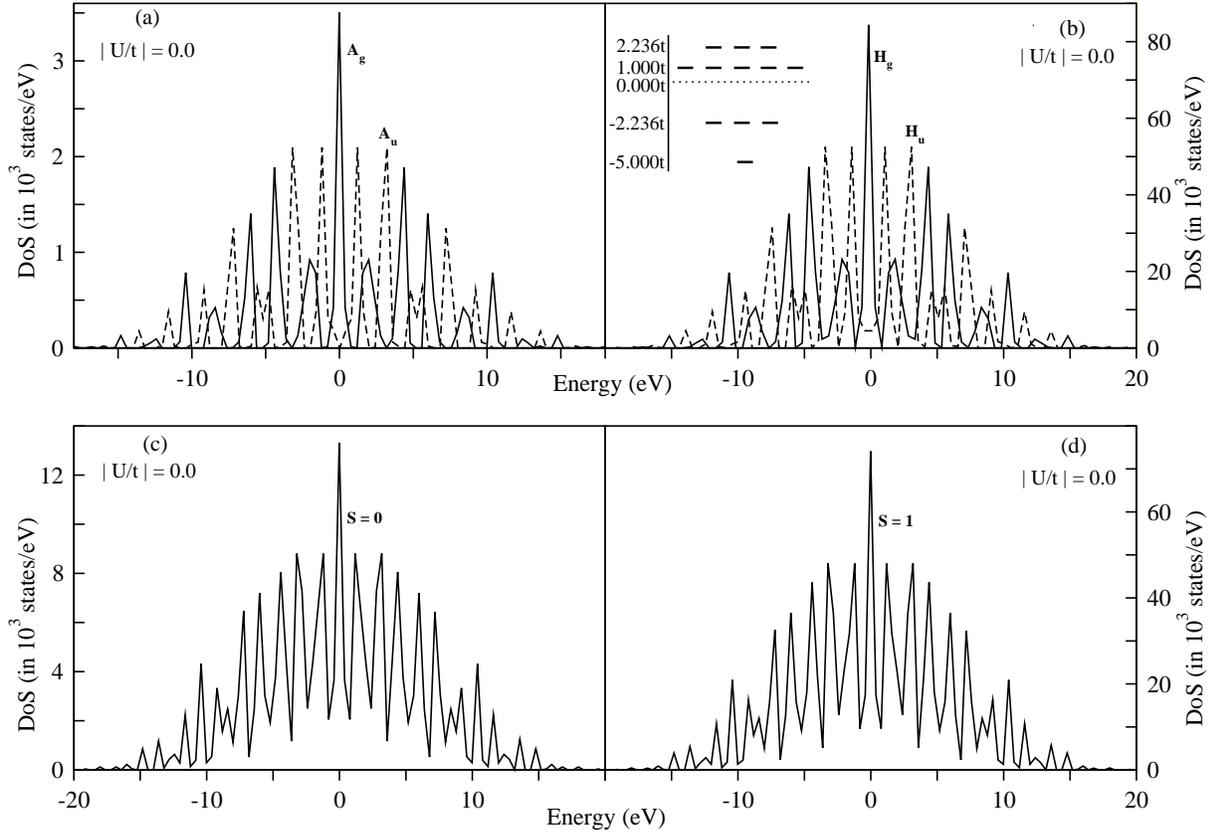}}
\caption{{\small DoS profiles for (a) A$_g$ and A$_u$ and (b) H$_g$ and H$_u$ 
spaces. For other symmetry subspaces, the DoS profiles are similar. We note 
the difference in profiles for $g$ and $u$ spaces. DoS profiles for (c) S = 0 
and (d) S = 1 are also given. Inset of (b) gives the one-particle spectrum for 
the regular icosahedron in the H\"uckel model.}}
\label{fig3}
\end{figure}

\begin{figure}[t]
{\includegraphics[width=1.0\textwidth]{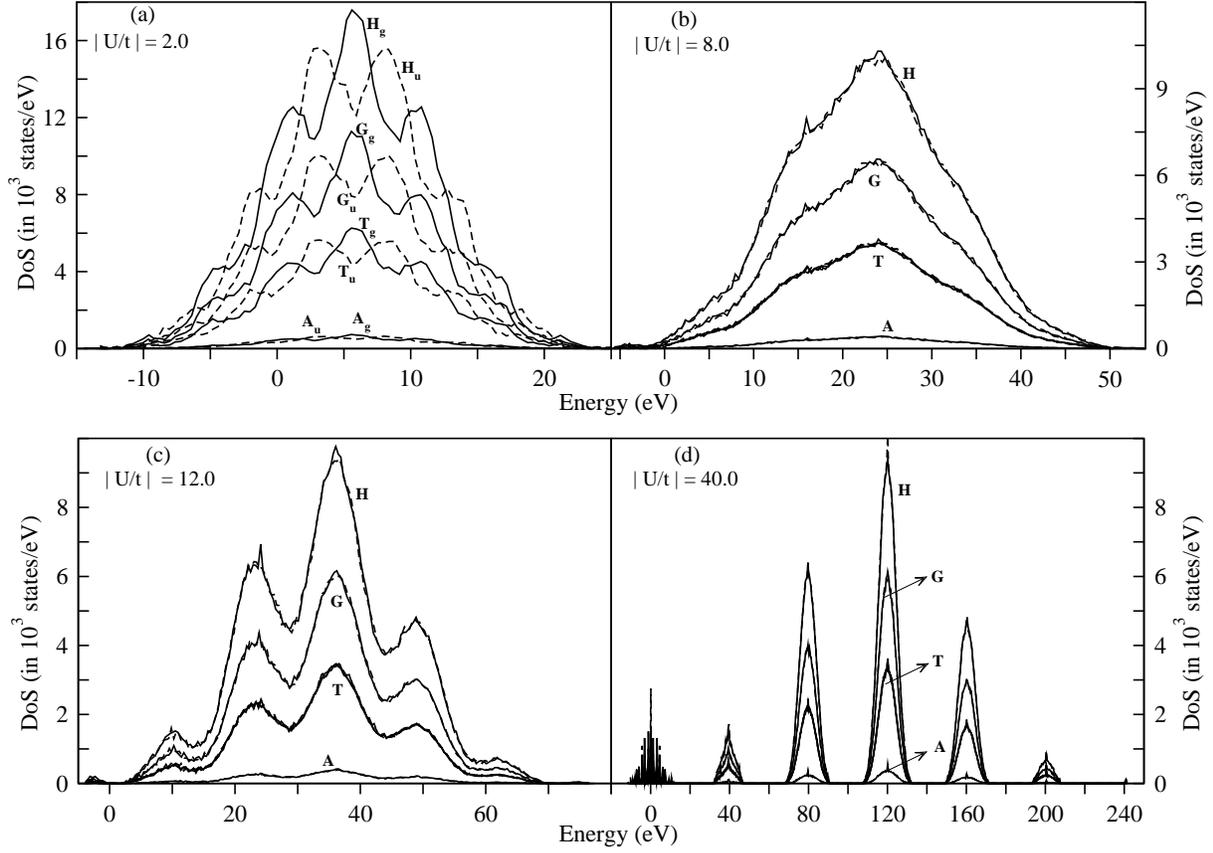}}
\caption{{\small DoS profiles in various symmetry subspaces for four different 
$|U/t|$ values are shown. Note, in (a), profiles for T$_{1g}$ and T$_{2g}$ 
coincide and simply referred to as T$_g$. Same is for the triply degenerate $u$ 
space. In (b), (c) and (d), DoS profiles for corresponding $u$ and $g$ spaces 
coincide, so they are referred to by their common irreducible representation 
symbols.}}
\label{fig4}
\end{figure}

\begin{figure}[t]
{\includegraphics[width=1.0\textwidth]{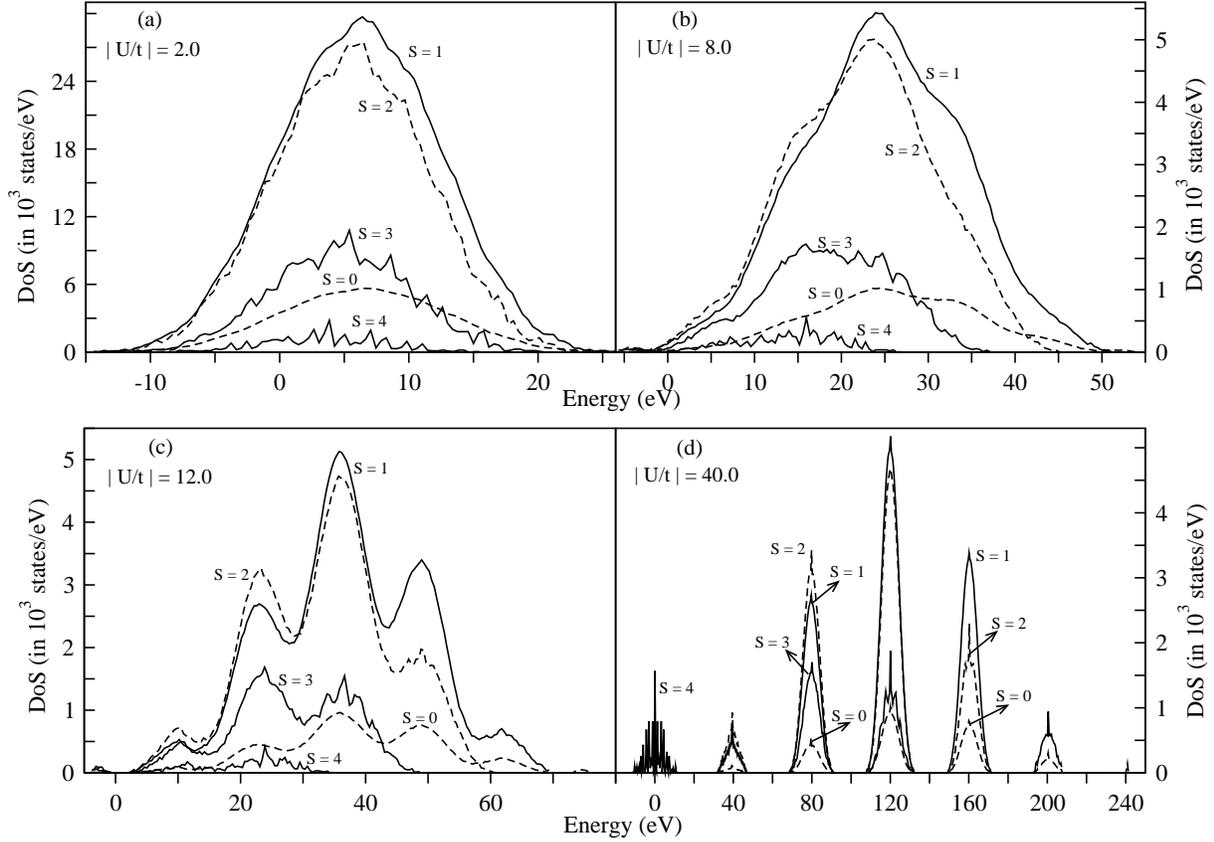}}
\caption{{\small DoS profiles of various spin subspaces for four different 
$|U/t|$ values are shown.}} 
\label{fig5}
\end{figure}

In Fig. (\ref{fig3}), we show the many-body DoS for the H\"uckel model in the 
the various $A_g,~H_g$ and $A_u,~H_u$ spaces. Here, we have summed over all 
spin states, for simplicity. We have also shown the DoS plots of different 
spin space, in which they are summed over all the irreducible spaces. Two 
things are worth noting. Firstly, the DoS displays a symmetry about zero of 
energy in each of the subspaces, even though the system does not possess the 
e-h symmetry. In fact, it is clear from the one particle spectrum that there 
is no symmetry in the one-particle energy levels about zero energy. Secondly, 
the DoS profile of the $g$ subspace shows peaks where ever there is a valley 
in the DoS profile of the $u$ subspace. This is indeed true also for other 
irreducible representations not shown in the figure. The reason for this 
symmetry in the DoS plots is because the sum of the one 
particle eigenvalues are zero and follows from the fact that in H\"uckel 
model, with all site energies set to zero, the diagonal matrix elements are 
all zero. This implies $\sum_i 2 \epsilon_i~ =~0$, where $\epsilon_i$ are the 
molecular orbitals (MO) energies. Thus, for any given occupancy pattern of the 
MOs at half-filling, we have $\sum_i n_i \epsilon_i~ =~ - \sum_i (2-n_i) 
\epsilon_i$ and since $\sum_i n_i = \sum_i (2-n_i)$, at half filling, we find 
that for every many-body state of energy $E_k$ there exists a many-body state 
of energy $-E_k$, even though the MO energies $\epsilon_i$ do not satisfy the 
pairing theorem \cite{coulson}. Thus, the symmetry in DoS plots is not a 
consequence of the pairing theorem but due to the fact that the magnitude of 
the sum of the energies of the bonding MOs is equal to the magnitude of the 
sum of the energies of anti-bonding MOs. This is in fact a general result for 
the H\"uckel model with 
equivalent sites. The second observation about the location of valleys and 
peaks in the DoS of the states with  $g$ and $u$ symmetries is due to the fact 
that the molecular orbital occupancies which give the $g$ and $u$ 
representations are different due to symmetry considerations.
The DoS plots in various spin subspaces are also shown in Fig. (\ref{fig3}). 
We find that they show several peaks in each total spin space. We show in 
Fig. (\ref{fig4}) DoS profiles for the Hubbard model in various symmetry 
subspaces for different values of $|U/t|$ and in Fig. (\ref{fig5}) we show the 
same in different total spin spaces. The evolution of DoS with correlation 
strength is interesting. Firstly, we 
note that for $|U/t| =2.0$, the sharp peaks in the DoS found in the H\"uckel 
model are broadened. The peaks in the $g$ subspace coincide with the troughs 
in the $u$ symmetry and {\it vice versa}. The ground state energy in the 
presence of correlations is higher than in the H\"uckel model, as expected.

In the very strong correlation regime, ($|U/t| = 12$ and $|U/t| = 40$ we again 
find peaks in the DoS in all the subspaces. What is interesting is that peaks 
appear at almost the same value of energy in $g$ and $u$ subspaces, unlike in 
the non-interacting or weakly interacting model, and are approximately $|U/t|$ 
apart in energy. This can be understood by noting that the many body space can 
be subdivided into space of all singly occupied sites; space of one empty, one 
doubly occupied and rest singly occupied sites; space of two empty, two doubly 
occupied and rest singly occupied sites and so on. The interaction energies of 
these class of states is 0, $U$, $2 U$, etc. The transfer term leads to 
weak admixture of these states, and in the strong correlation limit results in
broadening of the DoS peaks centered at the energies 0, $U$, $2 U$ etc. Thus, 
the DoS plots, although look similar in both the small $|U/t|$ and the large 
$|U/t|$ limits, their origin as well as their location is different. It is also 
worth noting that the DoS curves centered at different energies are similar 
for all the subspaces.  However, in the large $|U/t|$ limit, the DoS curves 
for the same total spin centered around different energies are not similar, 
showing that we do not have a strict spin-charge separation in icosahedron in 
this limit. For, if indeed we had such a separation, we would expect very 
similar DoS for the same total spin, for different number of doubly occupied 
sites, when the allowed number of total spin states is large.

For $|U/t| = 8.0$, the DoS is very different.  We find that there is a single 
broad peak and all the other peaks are small inflexions superimposed over the 
peak. The "band width" of the one-particle spectrum (see Fig. \ref{fig3}b, 
inset) is $7.236 t$ 
and the Hubbard correlation strength is very close to this value. Thus, for an 
icosahedral cluster, the parameters are at the intermediate correlation regime 
and leads to a smearing of the structure which is found at the weak and strong 
correlation limits. However, independent of the correlation strength, we find 
that the DoS curves are nearly identical for the T$_{1g}$ and T$_{2g}$ spaces 
and also for the T$_{1u}$ and T$_{2u}$ spaces.

\begin{table}[t]
\caption{{\small Lowest and second lowest energies for each symmetrized
spin sector (for PPP model). All energies are in eV.}}
\begin{center}
\begin{tabular}{|l||c|c|c|c|c|c|c|} \hline
S$_{tot} \rightarrow$ & 0 & 1 & 2 & 3 & 4 & 5 & 6 \\
$\Gamma$ $\downarrow$ &   &   &   &   &   &   &  \\ \hline \hline
A$_{g}$ & 0.000 & 8.154 & 5.600 & 9.976 & 16.871 & 40.390 & 36.724 \\
        & 1.533 & 8.835 & 8.618 & 15.601 & 26.400 & 42.449 & $-$    \\
[.5ex] \hline
T$_{1g}$ & 0.912 & 0.672 & 5.918 & 9.011 & 18.255 & 40.390 & $-$ \\
         & 6.660 & 1.249 & 8.726 & 10.594 & 18.620 & 42.449 & $-$ \\
[.5ex] \hline
T$_{2g}$ & 0.830 & 0.665 & 5.878 & 9.042 & 18.577 & 40.390 & $-$ \\
         & 6.228 & 1.229 & 9.007 & 10.171 & 19.058 & 42.449 & $-$ \\
[.5ex] \hline
G$_{g}$ & 0.058 & 0.134 & 5.485 & 8.631 & 17.626 & 30.589 & $-$ \\
        & 1.178 & 0.638 & 6.891 & 9.015 & 18.281 & 41.322 & $-$  \\
[.5ex] \hline
H$_{g}$ & 0.777 & 0.122 & 0.279 & 8.895 & 16.967 & 26.206 & $-$  \\
        & 0.831 & 1.393 & 5.401 & 9.160 & 17.609 & 30.428 & $-$   \\
[.5ex] \hline
A$_{u}$ & 2.339 & 3.502 & 2.713 & 12.029 & 21.755 & $-$ & $-$ \\
        & 3.725 & 4.274 & 7.900 & 16.492 & 26.126 & $-$ & $-$  \\
[.5ex] \hline
T$_{1u}$ & 3.846 & 2.349 & 2.937 & 6.888 & 20.761 & 31.775 & $-$ \\
         & 4.051 & 2.844 & 6.943 & 11.175 & 24.184 & 35.225 & $-$ \\
[.5ex] \hline
T$_{2u}$ & 2.653 & 2.382 & 2.854 & 10.950 & 21.087 & 23.484 & $-$ \\
         & 4.111 & 2.773 & 7.506 & 11.929 & 24.249 & 32.728 & $-$  \\
[.5ex] \hline
G$_{u}$ & 2.724 & 2.233 & 2.630 & 10.982 & 15.232 & 33.555 & $-$ \\
        & 3.461 & 2.690 & 4.005 & 11.857 & 20.867 & 38.355 & $-$  \\
[.5ex] \hline
H$_{u}$ & 2.225 & 2.372 & 2.414 & 11.638 & 15.747 & 33.469 & $-$ \\
        & 2.795 & 2.676 & 2.689 & 11.881 & 20.653 & 38.586 & $-$ \\
[.5ex] \hline
\end{tabular}
\end{center}
\label{tbl3}
\end{table}

\subsection{PPP model studies}

\begin{figure}[t]
{\includegraphics[width=1.0\textwidth]{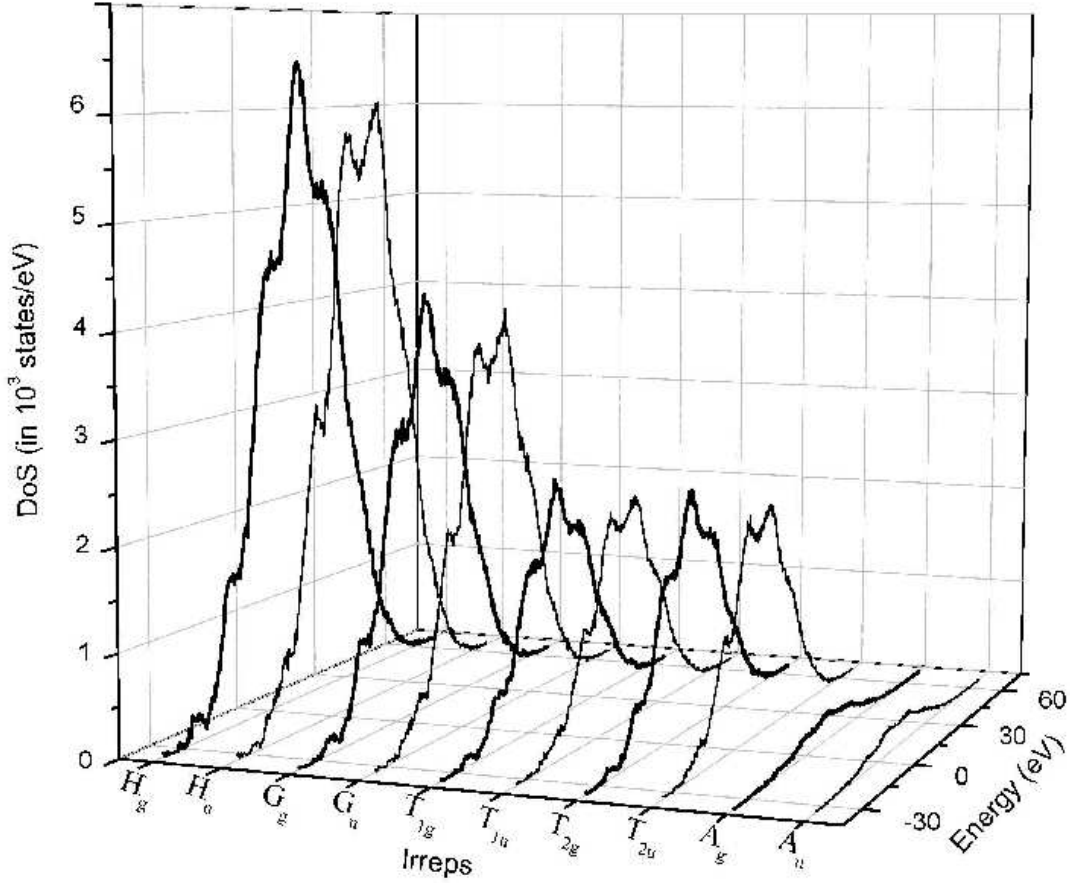}}
\caption{{\small DoS profiles of various symmetry subspaces for PPP model.}} 
\label{fig6}
\end{figure}

\begin{figure}[t]
{\includegraphics[width=1.0\textwidth]{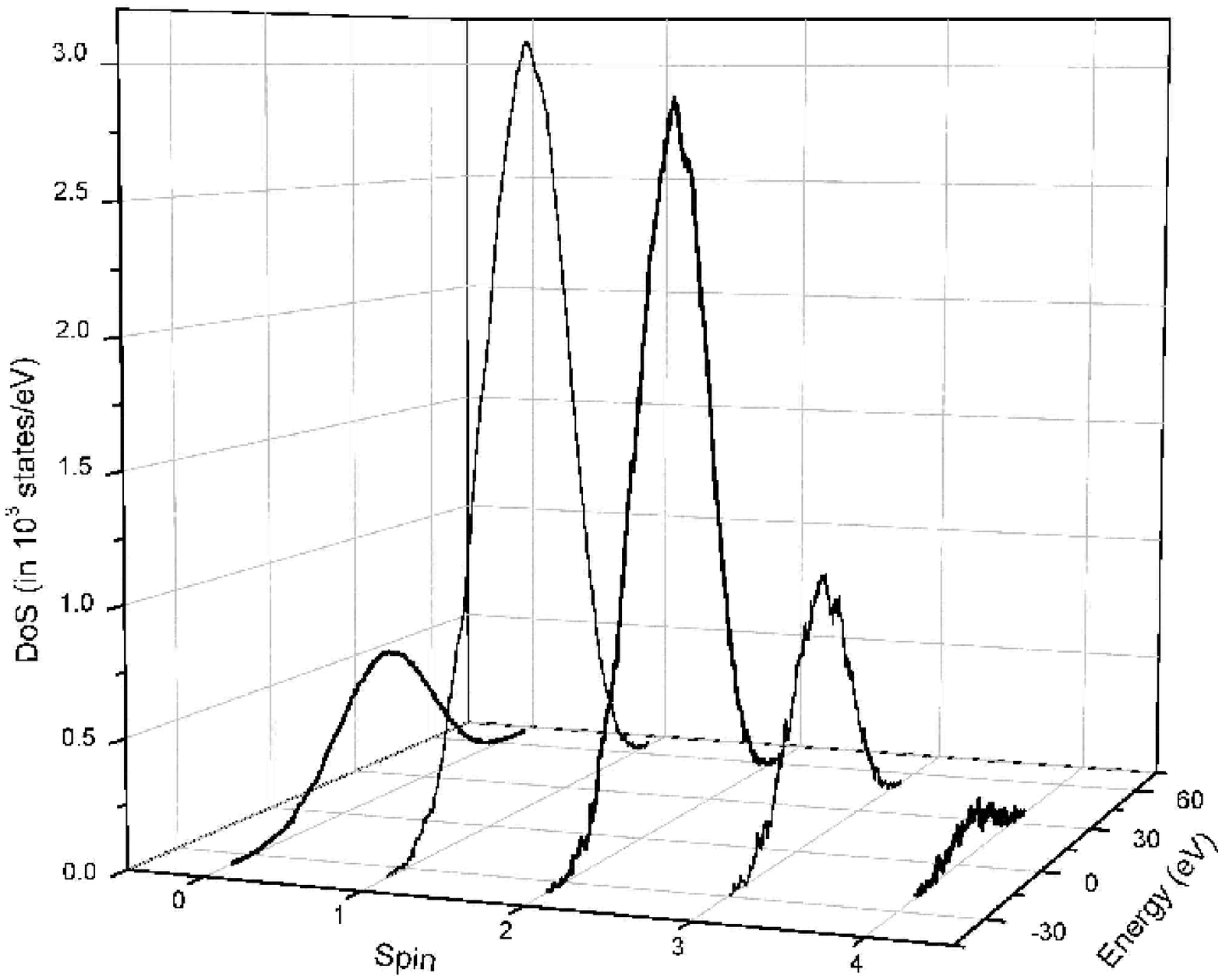}}
\caption{{\small DoS profiles of various spin subspaces for PPP model.}} 
\label{fig7}
\end{figure}

The Hubbard model is not the appropriate model for studying carbon systems as 
it neglects long-range interactions. The appropriate model for studying such 
systems is the PPP model, which we have employed for studying the cluster.

In Table (\ref{tbl3}), we show two lowest energy levels in each of the 
subspaces. We note that the ground state is the lowest energy state in the 
$A_g$ subspace with total spin zero. The one-photon gap is given by the lowest 
energy excitation to the $T_{1u}$ space for an Icosahedron. Thus, we find that 
lowest excitation gap is at energy of 3.846 eV. This can be compared with the 
excitation gap of 3.552 eV for a PPP chain of 12 carbon atoms \cite{soosppp}. 
We do not compare the excitation gaps to PPP ring of 12 carbon atoms, as the 
$N = 4n$ ($n$ integer) have very strong finite size effects due to degenerate 
partly filled highest occupied molecular orbitals. The second allowed 
excitation is at an energy of 4.051 eV. The two photon gap is to the second 
lowest energy state in the $A_g$ representation and is found to be 1.533 eV 
which is very low compared to a polyene chain ($\sim$ 3.0 eV). 

The lowest energy spin gap, from the singlet ground state is to the lowest 
energy spin 1 state in the $H_g$ space. This gap is 0.122 eV which is very low 
compared to the polyenes. In general, the singlet-triplet gaps in conjugated 
systems is about 60\% the optical gap and icosahedron seems to be an 
exception. Since singlet-triplet gap here is much higher compared to room 
temperature enery (about 0.025 eV), the system would show diamagnetic behavior 
below this temperature. There is also another triplet state which is about 
0.012 eV above the lowest energy triplet state. These observations imply that 
the icosahedral cluster would exhibit paramagnetism above room temperature due 
to significant population of these states. Based on the similar argument, we 
also conclude that the specific heat at low-temperature will be very small and 
increase exponentially with increasing temperature. The triplet-triplet (TT) 
excitation from the $H_g$ space is to states in $T_{1u},~ T_{2u},~ G_u ~ \rm 
{and} ~ H_u$ while from the $G_g$ state is to states $T_{2u}$, $G_u$ and 
$H_u$. This means that we would have a band of TT excitations starting from 
2.11 eV. Regarding higher spin excitations, there is a low energy quintet 
state about 0.279 eV above the ground state and a few other quintet 
excitations of energies between 2.414 and 2.937 eV. All other spin excitations 
are very high energy excitations, as the higher spin states have very low 
kinetic stabilization.

In Figs. (\ref{fig6}) and (\ref{fig7}), we show the density of states plots 
for different symmetry 
subspaces and different total spins. We note from the figures that the 
icosahedral cluster of conjugated Carbon atoms belongs to the weakly 
correlated regime since the DoS peaks in the $g$ and $u$ spaces do not 
coincide in energy. We also find that this conclusion is corroborated by the 
DoS plots for different total spin states. The $S=0$ DoS plot shows a 
featureless broad peak, as seen for small $|U/t|$ values of the Hubbard model. 
The higher spin states also show broad peaks consistent with the weak 
correlation regime. Even in the PPP model, the DoS plots for the $T_{1g}$ and 
$T_{2g}$ are very similar and same is the case with the $T_{1u}$ and $T_{2u}$ 
states. These DoS plots would also indicate the nature of the electronic 
spectra in these systems.

One of the most fascinating molecules to have been discovered is C$_{60}$, 
which also has icosahedral symmetry. The H\"uckel band width of C$_{60}$ is 
5.618 $t$, when the transfer integrals for the hexagon-pentagon and the 
hexagon-hexagon bonds are taken to be the same. This is much smaller than the 
7.236 $t$ found for icosahedron. This is largely due to the different number 
of bonds per site (2.5 bonds / site for icosahedron compared to 1.5 bonds / 
site for C$_{60}$) in the two systems. For this reason, we expect PPP model 
with standard parameters of C$_{60}$ to be in a more strongly correlated 
regime than the icosahedron. This should also reflect in the electronic 
spectra of C$_{60}$. The full spectrum of icosahedron will also be helpful in 
gaining insights into the contributions of different states to the linear and 
nonlinear optical response of the system.

\section{Summary}

In this paper, we have considered the long standing problem of both spacial 
and spin symmetry adaptation for arbitrary point groups. We have shown that by 
using the strengths of the VB and the constant $M_S$ methods, we can have a 
hybrid scheme which exploits the full symmetry of a non-relativistic 
Hamiltonian. We have illustrated this by applying to the nontrivial case of an 
icosahedral cluster.  We have obtained all the eigenstates of the cluster by 
our method. The hybrid method is less demanding on both memory and {\it CPU} 
time of a computer and is easy to implement. We have obtained the DoS of the 
Hubbard model for different Hubbard parameters $U$ and of the PPP model for an 
icosahedral cluster. These plots show different characteristics as a function 
of interaction strength in the Hubbard model. The PPP model studies indicate 
that while the one-photon gaps are comparable with other conjugated systems, 
the spin gaps are unusually small. This may lead to significant population of 
the magnetic states at room temperature. These studies have a bearing on the 
C$_{60}$ system which also possesses icosahedral symmetry. The method 
discussed here will be of considerable importance in studying the dynamics and 
finite temperature properties of systems whose Hamiltonians are amenable to 
exact diagonalization. While we have illustrated the method using a highly 
symmetric Hamiltonian, the method is very general and applicable to systems 
belonging to any point group. In point groups with lower symmetry, while the 
advantage of automatically labeling the states exists, the actual savings in 
computational effort would be decreased. 

\section{Acknowledgments}
This work was supported by Department of Science and Technology, Government of 
India, through various projects. We are also pleased to acknowledge Prof. 
Diptiman Sen for valuable discussions.

\end{document}